\documentclass[pra,twocolumn,showpacs,preprintnumbers,amsmath,amssymb]{revtex4-1}
\usepackage{graphicx}
\newcommand*\Diff[1]{\mathop{}\!\mathrm{d^#1}}
\newcommand*\erfc{\text{erfc}}
\newcommand*\re[1]{\text{Re}\left(#1\right)}
\newcommand*\im[1]{\text{Im}\left(#1\right)}

\DeclareGraphicsExtensions{.eps}
\begin{document}

\author{Matthew Thornton}
\email{mt45@st-andrews.ac.uk}
\affiliation{University of St Andrews}

\author{Hamish Scott}
\affiliation{University of St Andrews}

\author{Callum Croal}
\affiliation{University of St Andrews}

\author{Natalia Korolkova}
\affiliation{University of St Andrews}

\title{Continuous-variable quantum digital signatures over insecure channels}
\begin{abstract}
Digital signatures ensure the integrity of a classical message and the
authenticity of its sender. Despite their far-reaching use in modern
communication, currently used signature schemes rely on computational
assumptions and will be rendered insecure by a quantum computer. We present a
quantum digital signatures (QDS) scheme whose security is instead based on the
impossibility of perfectly and deterministically distinguishing between quantum
states. Our continuous-variable (CV) scheme relies on phase measurement of a
distributed alphabet of coherent states, and allows for secure message
authentication against a quantum adversary performing collective beamsplitter
and entangling-cloner attacks. Crucially, for the first time in the CV setting
we allow for an eavesdropper on the quantum channels and yet retain shorter
signature lengths than previous protocols with no eavesdropper. This opens up
the possibility to implement CV QDS alongside existing CV quantum key
distribution (QKD) platforms with minimal modification.
\end{abstract}
\maketitle

\section{Introduction}\label{sec:introduction}
Digital signatures are among the most commonly used primitives in modern
cryptography \cite{Diffie1976, Schneier1996}. Like a handwritten signature, a
digital signature ensures the authenticity of both a classical message and its
sender. Inherently involving multiple parties, the signature prevents a
malevolent party from creating a false message and attributing it to an honest
party; and if the signature convinces one party that a message is genuine then
it should convince the other parties. Despite their far-reaching use in our
modern technological infrastructure--e.g. in e-commerce, online banking and
checking the integrity of downloads--currently used signature schemes such as
RSA, DSA and ECDSA will be rendered insecure by a future quantum computer
\cite{Shor1997, Amiri2015, Nielsen2010, Schneier1996}.
\par
Quantum digital signatures (QDS) rectify this by basing their security not on
computational assumptions about difficult to invert ``one-way functions", but on
physical properties, namely, the impossibility to perfectly and
deterministically distinguish between non-orthogonal quantum states
\cite{Gottesman2001, Amiri2015, Nielsen2010, Wallden2015, Donaldson2016,
Dunjko2014, Collins2014, Croal2016, Andersson2006, Amiri2016, Collins2016,
Puthoor2016, Yin2016}. Quantum states are used to distribute a classical
signature which is later used to sign a classical message. While modern QDS
shares some similarities in implementation with quantum key distribution (QKD),
their aims differ significantly. The most notable difference is that
that--unlike QKD--in QDS any subset of the participants may be dishonest, and
each dishonest player may have different goals and strategies which must be
considered in a full proof.
\par
Although the first QDS schemes \cite{Gottesman2001} relied on infeasible
requirements like the production and storage of large entangled states, there
has since been a push towards practical and implementable QDS. The past decade
has done away with the need for an optical multiport \cite{Wallden2015,
Donaldson2016}, quantum memory \cite{Dunjko2014, Collins2014}, and recent progress has even removed the need for single
photon sources \cite{Croal2016, Andersson2006, Amiri2016, Collins2016} for secure QDS. More recently the assumption of secure quantum channels
for distribution of quantum states has been discarded, and modern QDS protocols
take into account both the ability for a player inside the scheme to be
dishonest, and the presence of an external eavesdropper \cite{Amiri2016,
Puthoor2016, Yin2016}.
\par
As with QKD, there are two approaches to QDS, the discrete-variable (DV) and
continuous-variable (CV) protocols. DV QDS relies on photon-number detection of
either weak coherent pulses \cite{Amiri2016, Collins2016, Collins2017,
Roberts2017, Zhang2018} or single photons \cite{Yin2016, Yin2017a}. The
relatively low dimensional Hilbert space required for these schemes allows for
an advanced level of security analysis, and their high resilience to loss allows
for long distances to be bridged securely, with $O\left(10^{-1}\right)$~seconds
required to sign a 1-bit message \cite{Collins2017, Collins2018, Yin2017}. The
first QDS scheme over insecure channels required a signature length of
$7.7\times10^5$ to sign a $1$~bit message, in a scheme similar to decoy-state
QKD \cite{Amiri2016}.
\par
In contrast, CV QDS encodes information into continuous degrees of freedom,
usually the phase of the electromagnetic field, and homodyne detection
\cite{Croal2016}--though a ``hybrid'' scheme has been proposed
\cite{Andersson2006} and implemented \cite{Clarke2012}, relying on both
phase-encoded coherent states and single-photon detection. Despite the
theoretical difficulties in dealing with large Hilbert spaces in a cryptographic
setting the CV platform is much easier to implement, operates at room
temperature and can use standard telecom hardware, making it thus closer to
currently implemented large-scale infrastructure
\cite{Weedbrook2012,Papanastasiou2018}.
\par
In the present paper we introduce a new CV QDS protocol based on a
discrete-modulated alphabet of coherent states, and the heterodyne detection of
phase. Such cheap and readily available resources make our scheme highly
compatible with telecom infrastructure. Crucially, and in contrast to our
previous paper \cite{Croal2016}, we now take into account the fact that the
quantum distribution channels may in general be insecure and under the control
of a malevolent party. Thus, we guard not only against dishonest participants
inside the protocol but also against an external eavesdropper.
\par
Our scheme is the first fully CV QDS scheme to run over insecure quantum
channels. Remarkably, despite relaxing an assumption on the quantum channels we
are able to reduce the number of quantum states required to securely sign a
message. We provide a conceptually new security proof and demonstrate that the
success probability of an eavesdropper can be made arbitrarily small. Our
security proof provides collective security against both beamsplitter attacks
and entangling-cloner attacks, with the main feature of the proof being that a
dishonest player may fail to correctly identify an element of the signature and
yet still remain undetected to the honest parties.
\par
The paper is structured as follows. In section~\ref{sec:protocol} we describe
our protocol and briefly discuss the origin of its security. Our security proof
follows in section~\ref{sec:secproof}, and in section~\ref{sec:performance} we
analyse the protocol's performance. Finally, in section~\ref{sec:discussion} we
compare our protocol to its nearest competitors and discuss potential extensions
to our security analysis. Technical details and a generalisation of the protocol
may be found in the appendices.

\section{Protocol description}\label{sec:protocol}
In the simplest instance we consider a signature scheme involving three parties:
a sender, Alice ($A$), and recipients, Bob ($B$) and Charlie ($C$). Alice wishes
to send a classical $1$~bit message $m$ to Bob, which he will forward to
Charlie. This scheme may be readily extended to include more players
\cite{Arrazola2015} or longer messages \cite{Wang2015, Wang2017}.
\par
In a successful QDS scheme, Bob and Charlie should be able to determine that
Alice is the genuine author of $m$. In particular, the scheme should guard
against a dishonest player--or an external Eve--from successfully forging a
message which is then accepted as genuine. It should also prevent Alice from
\emph{repudiating}, which occurs if Alice convinces Bob that a message is
genuine and Charlie that it is fake. The scheme should succeed if all parties
are honest, except with negligible probability. We allow at most one of the
players to be dishonest, noting that a three-party protocol fails trivially if
more dishonest players are permitted.
\par
We focus on the quadrature phase-shift keying (QPSK) alphabet of $4$
phase-encoded coherent states, denoted $\mathcal{A}_4$ \cite{Papanastasiou2018,
Weedbrook2012, Leverrier2011} distributed equally around the axis in phase space
(Fig.~\ref{fig:1}, inset). In Appendix~\ref{appendix:larger} we demonstrate how
larger alphabets $\mathcal{A}_N$ may be incorporated into our proof and
analysis.
\par
The QDS scheme is split into two stages, Distribution and Messaging, which can
occur with significant time delay. The coherent states are sent by Alice and
measured by Bob and Charlie during Distribution, while during Messaging Alice
will send the message $m$ to Bob, accompanied by a classical signature. This
signature is her classical declaration of which quantum states she sent. By
comparing her declaration to their measurement outcomes, Bob and Charlie can
determine whether $m$ is genuine. Our protocol is outlined in Fig.~\ref{fig:1}
and described in full, below.

\begin{figure}[htp]
\centering
\includegraphics[width=\linewidth]{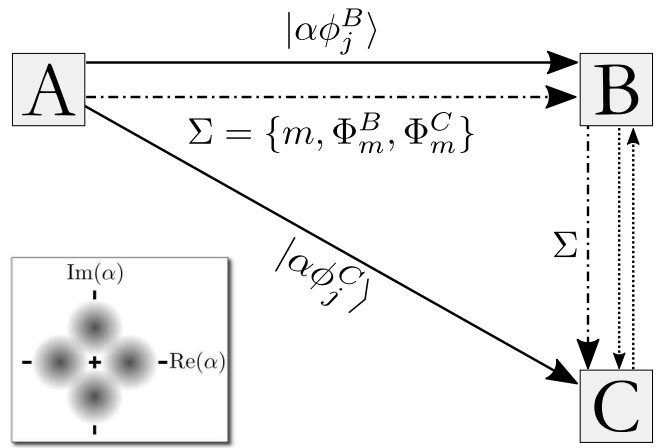}
\caption{\label{fig:1}Schematic of $3$~party QDS. The parties share quantum
distribution channels (solid lines), public classical channels (dot-dashed
lines), and Bob and Charlie share an encrypted classical channel (dashed lines).
Initially, sender $A$ distributes her classical signatures $\{\Phi_m^B,
\Phi_m^C\}$ via the quantum states $|\alpha\phi_j^{B,C}\rangle$ by encoding into
the QPSK alphabet. Then she sends a message $m$ to recipients Bob ($B$) and
Charlie ($C$), with the corresponding signatures, through the public classical
channel. $B$ and $C$ use the signatures to authenticate $m$. The encrypted
classical channel is used during the Symmetrization step of the protocol. Inset:
the QPSK alphabet of coherent states with amplitudes $\alpha \in \mathbb{C}$.}
\end{figure}
\par
\subsection{Distribution stage: $1-4$}
\noindent $1.$ Alice wishes to send a signed $1$~bit message $m$ to Bob and
Charlie. For each possible $m$, Alice creates two different classical strings,
one for Bob and one for Charlie, $\Phi_m^{\left(B,C\right)} =
\{\phi_j^{\left(B,C\right)}\}_{j=1}^{L}$ where the $\phi_j$ are phases chosen
uniformly at random from our alphabet $\mathcal{A}_4 = \{1, i , -1, -i \}$. The
signature length $L$ is an integer suitably chosen to ensure security.
\par
\noindent $2.$ For each element $\phi_j^{\left(B,C\right)}$ Alice forms the
corresponding coherent state $| \alpha \phi_j^{\left(B,C\right)}\rangle$ and
sends it to $B,C$, Fig.~\ref{fig:1}. The amplitude $\alpha$ is chosen to
optimise security. By analogy with classical digital signatures, we may think of
the $\Phi_m^{\left(B,C\right)}$ as Alice's private keys, and the corresponding
sequences of quantum states as her public keys. In contrast to our previous QDS
protocol we take $\Phi_m^B \ne \Phi_m^C$ \cite{Croal2016, Amiri2016}. Since
coherent states are non-orthogonal an eavesdropper on the quantum channel cannot
gain full information about Alice's signatures.
\par
\begin{figure}[htp]
\centering
\includegraphics[width=\linewidth]{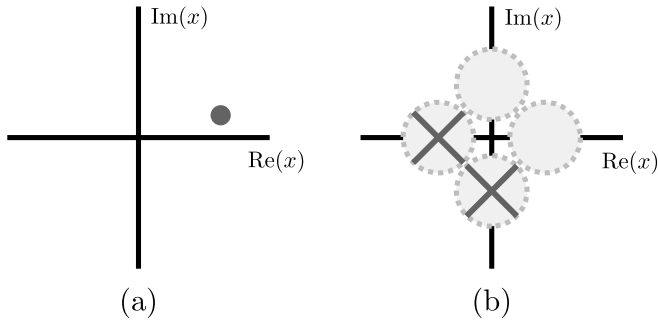}
\caption{\label{fig:2}After measuring the phase of a distributed coherent state,
Bob and Charlie eliminate the two alphabet states which were least likely to
have given that outcome. (a) An individual measurement outcome $x$ with $\re{x}
\ge 0$, $\im{x} \ge 0$. (b) The corresponding eliminated signature element. The
states $|-\alpha\rangle, |-i\alpha\rangle$ are the least likely from our
alphabet to give $x$, so they are eliminated.}
\end{figure}

\noindent $3.$ Bob and Charlie measure the phases of the received states by
heterodyne detection \cite{Weedbrook2012}, and keep a record of the alphabet
states which are most incompatible with their measurements, Fig.~\ref{fig:2}.
For example, if Bob measures $b \in \mathbb{C}$ with $\re{b} >0$ and $\im{b}>0$
then he will ``eliminate'' states $|-\alpha\rangle$ and $|-i\alpha\rangle$ since
these are the least likely of Alice's sent states to generate this outcome.
Recipients Bob and Charlie each now possess an ``eliminated signature''
\cite{Wallden2015, Croal2016, Donaldson2016} of length $L$ containing a record
of which states were eliminated at each position in the sequence. Since
measurements are performed immediately on receipt of the states, no quantum
memory is required \cite{Wallden2015}.

\par
\noindent $4.$ \emph{Symmetrization:} Bob and Charlie swap a random $L/2$
elements of their eliminated signatures over their encrypted classical channel,
keeping the positions and values of the swapped elements secret from Alice.
Signature elements which are forwarded by a recipient will no longer be used by
them in the protocol. This swapping will provide security against repudiation.
Bob and Charlie each now possess an eliminated signature in two halves, those
elements received directly from Alice and those received during this
Symmetrization step.

\subsection{Messaging stage: $5-7$}
\noindent $5.$ Messaging can occur at any time after Distribution. To sign $m$,
Alice sends to Bob the classical triplet $\Sigma = \left(m, \Phi_m^B,
\Phi_m^C\right)$, consisting of the message $m$ and classical information
$\Phi_m^{B,C}$ about the correponding sent quantum states.
\par
\noindent $6.$ Bob compares elements of $\Phi_m^{\left(B,C\right)}$ with the
corresponding elements of his eliminated signatures, and counts the number of
mismatches. A mismatch occurs if Bob has eliminated a state which Alice claims
to have sent. Note that even when all parties are honest there will still be
some probability of mismatch as the alphabet states are not orthogonal, so if
Alice sent $|\alpha\rangle$, Bob or Charlie can still measure $\re{x}<0$. If
there are fewer than $s_B L/2$ mismatches, for threshold $s_B$, between each
half of his eliminated signature and Alice's declaration, then Bob accepts $m$
as genuine. Otherwise he rejects it and the protocol aborts.
\par
\noindent $7.$ Bob forwards $\Sigma$ to Charlie, who similarly checks for
mismatches. Charlie accepts the message if there are fewer than $s_C L/2$
mismatches between elements of $\Sigma$ and each half of his eliminated
signature. If Charlie also accepts $m$ then the protocol has succeeded,
otherwise it aborts. The crucial parameters in steps $6$ and $7$ are the
$s_{\left(B,C\right)}$, which can be chosen to optimize security.

\section{Security proof and attack analysis}\label{sec:secproof}
In order to be secure, a QDS protocol must abort when a forging or repudiation
attack is attempted. From steps $6$ and $7$ in the protocol, the protocol aborts
if the number of mismatches observed by Bob or Charlie is above $s_BL/2$ or
$s_CL/2$, respectively. We now demonstrate that a forgery or repudiaton attack
will induce such a large number of mismatches, so the attacks are therefore
detectable. We also show that in the absence of an attack the protocol succeeds,
that is, it is robust. The protocol fails if it allows a forging or repudiation
attack, or if it aborts even when all parties are honest. The proofs of
robustness and security against repudiation from \cite{Croal2016} can be
directly applied to our new protocol. For completeness we reproduce the key
results in Eqs.~(\ref{eqn:erob}),~(\ref{eqn:erep}) below, and full proofs may be
found in \cite{Croal2016}. Importantly, the security against forgery requires a
completely new analysis and this will be one of the main results presented in
the paper.

\subsection{Robustness}
A QDS protocol is called \emph{robust} if it succeeds when all parties are
honest, except with a negligible probability $\varepsilon_{rob}$. Since the
alphabet $\mathcal{A}_4$ of distributed states is highly non-orthogonal, even
when all parties are honest there is still a probability $p_{err}$ that a
recipient eliminates the state which Alice sent. However, $p_{err}$ is
predictable and can be estimated during the protocol. For a pure loss channel
with transmission $T$, the rate $p_{err}$ corresponds to the probability of a
heterodyne measurement outcome $\re{x} < 0$ when Alice sent the coherent state
$|\alpha\rangle$ with $\alpha \in \mathbb{R}\ge 0$
\begin{equation}\label{eqn:perr}
p_{err} = \frac{1}{2}\erfc\left(\sqrt{\frac{T}{2}}\alpha\right).
\end{equation}
With mismatch rate $p_{err}$, we may use Hoeffding's inequalities,
Appendix~\ref{appendix:hoeffding}, to bound the probability that Bob or Charlie
detect more than $s_{\left(B,C\right)}L/2$ mismatches as
\begin{equation}\label{eqn:erob}
\varepsilon_{rob} \le 2 \exp\left(-\left(s_{\left(B,C\right)} - p_{err}\right)^2
L \right)
\end{equation}
provided that $s_{\left(B,C\right)} > p_{err}$. Eq.~(\ref{eqn:erob}) is derived
in \cite{Croal2016} using Eq.~(\ref{eqn:ehoeff2}). The probability
$\varepsilon_{rob}$ of the protocol aborting even when all parties are honest
can thus be made arbitrarily small by choice of $L$.

\subsection{Security against repudiation}
Alice succeeds in a repudiation attack if she convinces Bob that a message is
genuine and Charlie that it is fake. During Messaging, Alice will declare
$\tilde{\Phi}_{m}^{B}$ and $\tilde{\Phi}_{m}^{C}$ chosen with the aim that there
should be fewer than $s_B L/2$ mismatches with each half of Bob's signature, but
more than $s_C L/2$ mismatches with at least one of Charlie's halves.
\par
Intuitively, security against repudiation arises from Symmetrization (step $4$
of the protocol). Since the swapping occured in secret from Alice, she does not
know who holds a particular eliminated signature element. Alice is therefore
unlikely to succeed in creating a declaration which will pass Bob's test but
fail Charlie's.
\par
The probability of successful repudiation is 
\begin{equation}\label{eqn:erep}
\varepsilon_{rep} \le 2 \exp\left(-\left(s_C - s_B\right)^2 \frac{L}{4}\right)
\end{equation}
provided that $s_C > s_B$. Eq.~(\ref{eqn:erep}) is derived fully in
\cite{Croal2016} using Eqs.~\ref{eqn:ehoeff}, \ref{eqn:ehoeff2}. The probability
$\varepsilon_{rep}$ of successful repudiation can thus be made arbitrarily small
by choice of $L$.

\subsection{Security against forgery}
It is the forging attack in which our analysis significantly differs from
\cite{Croal2016}. In a successful forging attack, a dishonest player will
declare some fake $m^\prime$ with the aim that it is accepted as genuinely
having originated with Alice. The message $m^\prime$ must have an appended
signature $\Phi_m^{\prime C}$, and so a forger's goal is to determine a fake
signature which will be accepted. Since Bob already knows half of Charlie's
eliminated signature elements--those which Bob himself forwarded--and since it
is easier to convince Charlie than Bob to accept a fake signature ($s_C > s_B$),
the most dangerous forger is a dishonest Bob. A bound for the probability that
Bob succeeds in a forging attack provides an automatic upper bound against any
other forging player.
\par
Since the empahsis of earlier papers \cite{Gottesman2001, Wallden2015,
Dunjko2014, Collins2014, Donaldson2016, Croal2016} was on internal dishonesty
between participants, eavesdropping on the quantum channels was not permitted. A
dishonest Bob had only his states received from Alice with which to gain
information about Charlie's outcomes. However, Alice distributed identical
signatures $\Phi_m^B = \Phi_m^C$ to Bob and to Charlie, so in-effect Bob had a
perfect copy of Charlie's signature. Now, to mitigate against an eavesdropping
Bob, we choose $\Phi_m^B \ne \Phi_m^C$, so that even though dishonest Bob can
gain some additional information on Charlie's quantum states, Bob now has a less
perfect copy of Charlie's states than he did previously \cite{Amiri2016}.
\par
Dishonest Bob will eavesdrop on the quantum states as Alice is distributing them
to Charlie, and will try to determine what he can declare in $\Phi_m^{\prime C}$
to not cause a mismatch. Defining $p_e$ as the probability that Bob will induce
a mismatch on a given signature element, and $s_C$ as Charlie's mismatch
threshold, the probability $\varepsilon_{forg}$ of a successful forging attack
is
\begin{equation}\label{eqn:eforg}
\varepsilon_{forg} \le 2\exp\left(-\left(p_e - s_C\right)^2 L\right)
\end{equation}
which is derived via Eq.~(\ref{eqn:ehoeff}) by analogy with \cite{Croal2016}.
Since $s_C \le p_e$ can be freely chosen, we must now calculate a lower bound on
the probability $p_e$.

\subsection{Calculating $p_e$}
Our main contribution is a conceptually new bound for $p_e$, which fully takes
into account the ambiguity in Bob's declaration. This ambiguity stems from the
following. Because Charlie eliminates two states, Fig.~\ref{fig:2}, there are
two possible states from $\mathcal{A}_4$ which Bob can declare without
introducing a mismatch. Therefore, the probability that Bob misidentifies an
element of the eliminated signature is not equivalent to the probability of
mismatch. We allow for this discrepancy by working directly in terms of mismatch
probability $p_e$ via an error variable $E$, and in the proof we highlight
quantities which are affected by the degeneracy in Bob's possible declaration.
In what follows we explicitly consider the QPSK alphabet $\mathcal{A}_4$. Our
security proof readily generalises to larger alphabets $\mathcal{A}_N$ with $N =
6, 8, 10, \dots$, and the required modifications to the proof are discussed in
Appendix~\ref{appendix:larger}.
\par
Let $X_j, 1 \le j \le L/2$ be an element of the half of $C$'s eliminated
signature which he received directly from Alice, and on which Bob will attempt
to gain some information. We write $X_j = \{x_1^j, x_2^j\}$ where $x_1^j$ and
$x_2^j$ describe the states from $\mathcal{A}_4$ which Charlie eliminated. The
$x_1^j$ and $x_2^j$ must be adjacent in $\mathcal{A}_4$, e.g. if $x_1^j = 1$
then $x_2^j = \pm i$. Let the string $Y = \{y_j\}_j$ be Bob's declaration,
subject to an unspecified but optimal POVM and classical strategy.
\par
A mismatch occurs when $y_j = x_1^j$ or $y_j = x_2^j$. To analyse the
probability that this occurs we define a variable $E_j$, which takes value $1$
if a mismatch occurs at position $j$ and $0$ otherwise. Then Bob's average
mismatch rate $p_e= P\left( E_j = 1 \right)$. Because $E_j$ can take one of two
values, the Shannon entropy $H\left(E_j\right)$ is equal to the binary entropy
$h\left(p_e\right) = - p_e \log p_e - \left(1-p_e\right)\log\left(1-p_e\right)$.
\par
Consider the conditional entropy $H\left(E_j, x_1^j, x_2^j | y_j\right)$, which
is related to the uncertainty about whether a mismatch has occured under Bob's
declaration $y_j$. Using the chain rule for conditional entropies
\cite{Nielsen2010} we write
\begin{equation}\label{eqn:chain1}
H\left(E_j, x_1^j, x_2^j | y_j\right) = H\left(E_j | x_1^j, x_2^j, y_j\right) +
H\left(x_1^j, x_2^j|y_j\right).
\end{equation}
Since a choice of $y_j, x_1^j$ and $x_2^j$ uniquely determines $E_j$, the first
term on the right hand side of Eq.~(\ref{eqn:chain1}) must equal $0$, so we have
\begin{equation}\label{eqn:chain2}
H\left(E_j, x_1^j, x_2^j | y_j\right) = H \left(x_1^j, x_2^j | y_j\right).
\end{equation}
Using the chain rule for conditional entropies once more on the left hand side
of Eq.~(\ref{eqn:chain1}),
\begin{align}\label{eqn:chain3}
H \left(E_j, x_1^j, x_2^j | y_j \right) &= H \left(x_1^j, x_2^j | E_j,
y_j\right) + H\left(E_j | y_j\right) \notag \\
&\le  H \left(x_1^j, x_2^j | E_j, y_j\right) + H\left(E_j\right)
\end{align}
where we can write the upper bound because conditioning cannot increase entropy.\par
Combining Eqs.~(\ref{eqn:chain2}),~(\ref{eqn:chain3}),
\begin{align}\label{eqn:chain4}
H\left(x_1^j, x_2^j | y_j\right) &\le H\left(x_1^j, x_2^j | E_j, y_j\right) +
H\left(E_j\right) \notag \\
& = \left(1-p_e\right) H \left(x_1^j, x_2^j | E = 0, y_j\right) \notag \\
& + p_e H \left(x_1^j, x_2^j | E_j = 1, y_j\right) + H\left(E_j\right).
\end{align}
Now because of the ambiguity in Bob's declaration, i.e. because there are two
eliminated signature elements consistent with a given $E_j = 0$ and $y_j$, and
since we can permute and relabel $x_1^j \leftrightarrow x_2^j$, we have
$H\left(x_1^j, x_2^j | E_j = 0, y_j\right) \le \log_2 4 = 2$. We also use the
fact that Charlie eliminates exactly half of the alphabet in order to write
$H\left(x_1^j,x_2^j|E_j=0,y_j\right) =H\left(x_1^j,x_2^j|E_j=1,y_j\right)$.
Therefore
\begin{equation}\label{eqn:chain5}
H \left(x_1^j, x_2^j | y_j\right) \le 2 + H\left(E_j\right) = 2 +
h\left(p_e\right).
\end{equation}
From the definition of mutual information \cite{Nielsen2010}, we have 
\begin{align}\label{eqn:mutinf}
H\left(x_1^j, x_2^j | y_j \right) &= H\left(x_1^j, x_2^j\right) - I\left(x_1^j,
x_2^j : y_j\right) \notag \\
&\ge 3 - \chi\left(x_1^j, x_2^j : y_j\right)
\end{align}
where we have used that the Holevo information $\chi$ maximizes the mutual
information $I$ over all POVMs, and that $H\left(x_1^j, x_2^j\right) = \log_2 8
= 3$ because of the four possible eliminated signature elements, and an
additional factor of $2$ due to relabeling.
\par
Combining Eqs.~(\ref{eqn:chain5}),~(\ref{eqn:mutinf}) we arrive at
\begin{equation}\label{eqn:hpe}
h\left(p_e\right) \ge 1 - \chi\left(x_1^j, x_2^j : y_j\right)
\end{equation} 
which is one of the main results of the paper. This inequality can be implicitly
solved for Bob's mismatch rate $p_e$, and provides security against collective
attacks provided that Bob's Holevo information $\chi$ can be estimated.

\subsection{Attack analysis}
Eqs.~(\ref{eqn:eforg}),~(\ref{eqn:hpe}) determine the required signature length
to provide security of our scheme in any situation where $\chi$ can be bounded.
In order to gain some insight into the behaviour of our protocol, in what
follows we restrict Bob to two classes of attack, the \emph{beamsplitter attack}
and the \emph{entangling-cloner attack}, which correspond to a pure-loss ($\xi =
0\%$) and thermal-loss ($\xi \ne 0\%$) channel, respectively. In an
implementation it is the measured excess noise $\xi$ at the receiver which will
determine the attack class that Bob is assumed to have performed
\cite{Papanastasiou2018}. In section~\ref{sec:discussion} we remark about the
optimality of these attacks.
\par
By definition Bob's Holevo information is \cite{Nielsen2010}
\begin{equation}\label{eqn:holevo}
\chi\left(x_1^j, x_2^j : y_j\right) = S\left(\rho_B^j\right) - \sum_{x_1^j,
x_2^j} p\left(x_1^j, x_2^j\right) S\left(\rho_B^{x_1^j, x_2^j}\right)
\end{equation}
with $S\left(.\right)$ the Von Neumann entropy, $\rho_B^j$ Bob's total \emph{a
priori} mixed state at position $j$, and $\rho_B^{x_1^j, x_2^j}$ Bob's state
conditioned on Charlie's $j^{\text{th}}$ eliminated signature element $X_j$
being $\{x_1^j, x_2^j\}$.

\subsubsection{Beamsplitter attack}
\begin{figure}[htp]
\centering
\includegraphics[width=\linewidth]{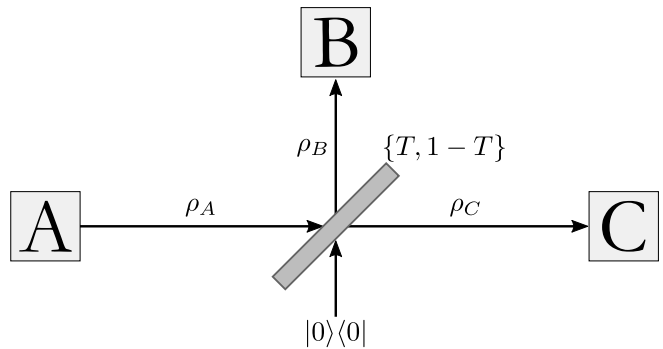}
\caption{Schematic of the beamsplitter attack. Alice distributes her state
$\rho_A$ through a lossy channel with transmission $T$, modelled as a
$\{T,1-T\}$~beamsplitter with vacuum input at the fourth port. Bob and Charlie
collect their states from the reflected and transmitted ports, respectively.}
\label{fig:3}
\end{figure}

We first consider the so-called beamsplitter attack, Fig.~\ref{fig:3}, in which
a purely lossy channel is modelled using a beamsplitter with transmission $T$
and vacuum input at the fourth port, and in which Bob collects his state
$\rho_B^j$ from the reflected output port.

Letting $|\alpha_k\rangle\langle\alpha_k|$ with $k = 1,2,3,4$ be an element of
$\mathcal{A}_4$, Alice's average input state may be written as

\begin{equation}\label{eqn:rhoA}
\rho_A^j = \frac{1}{4}\sum_{k=1}^{4} |\alpha_k\rangle\langle\alpha_k|
\end{equation}
which is transformed by the beamsplitter with vacuum input to become
\begin{align}
&\rho_{BC}^j = \notag \\
&\frac{1}{4}
\sum_{k=1}^4|\sqrt{T}\alpha_k\rangle_C\langle\sqrt{T}\alpha_k|\otimes|\sqrt{1-T}\alpha_k\rangle_B\langle\sqrt{1-T}\alpha_k|.
\end{align}
Then Bob's \emph{a priori} state $\rho_B^j$ is given by
\begin{equation}\label{eqn:rhoBbs}
\rho_B^j =
\frac{1}{4}\sum_{k=1}^4|\sqrt{1-T}\alpha_k\rangle_B\langle\sqrt{1-T}\alpha_k|
\end{equation}
from which the first term in Eq.~(\ref{eqn:holevo}) can be calculated.
\par
Charlie performs heterodyne measurement on his half of $\rho_{BC}^j$ and
receives outcome $c_j \in \mathbb{C}$. The state $\rho_{BC}^j$ is transformed as
\begin{align}\label{eqn:rhobcj}
&\rho_{B|c}^j \notag \\
&= \frac{1}{4} \sum_{k=1}^4 \langle c | \sqrt{T}\alpha_k\rangle \langle
\sqrt{T}\alpha_k | c \rangle |
\sqrt{1-T}\alpha_k\rangle_B\langle\sqrt{1-T}\alpha_k| \notag \\
&= \frac{1}{4}\sum_{k=1}^4 p\left(c | \alpha_k\right)
|\sqrt{1-T}\alpha_k\rangle_B\langle\sqrt{1-T}\alpha_k|
\end{align}
where $p\left(c|\alpha_k\right) = 1/\pi
\exp\left(-|c-\sqrt{T}\alpha_k|^2\right)$ is the probablity of Charlie measuring
$c$ when the state $|\sqrt{T}\alpha_k\rangle$ is received, and $|c\rangle$ is a
coherent state centered on $c \in \mathbb{C}$.
\par
\begin{figure}[htp]
\centering
\includegraphics[width=\linewidth]{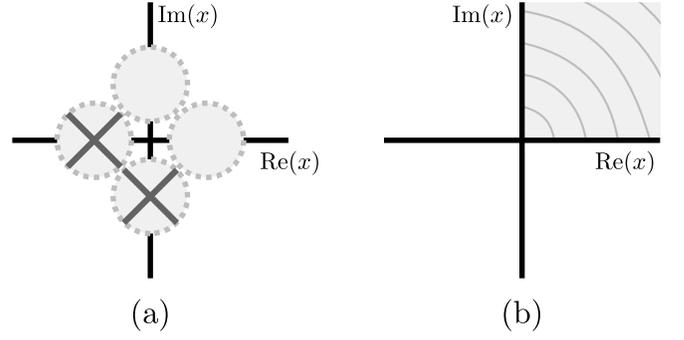}
\caption{A single eliminated signature element corresponds to an entire phase
quadrant as an allowed region for the phase measurement outcome. (a) An
eliminated signature element. (b) The element can be generated by any heterodyne
measurement outcome in the shaded region.}
\label{fig:4}
\end{figure}

On average, each eliminated signature element $X_j = \{x_1^j, x_2^j\}$ is
equally likely, so for Eq.~(\ref{eqn:holevo}) it will suffice to calculate
$S\left(\rho_B^{x_1^j,x_2^j}\right)$ for just one. An element $X_j$ is uniquely
determined by the quadrant in which the outcome $c$ lies, Fig.~\ref{fig:4}.
Using Eq.~(\ref{eqn:rhobcj}) we may write
\begin{equation}\label{eqn:int}
\rho_B^{x_1^j, x_2^j} = \int \rho_{B|c}^j \Diff2 c
\end{equation}
where the integration is perfomed over an entire quadrant in phase space. The
states $\rho_B^j$ and $\rho_B^{x_1^j, x_2^j}$ from
Eqs.~(\ref{eqn:rhoBbs}),~(\ref{eqn:int}) can be inserted into
Eq.~(\ref{eqn:holevo}) and the mismatch rate $p_e$ can now be calculated.

\subsubsection{Entangling cloner attack}\label{sec:ec}
\begin{figure}[htp]
\centering
\includegraphics[width=\linewidth]{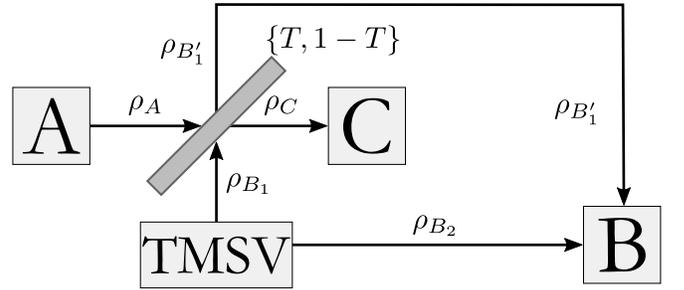}
\caption{Schematic of the entangling cloner attack. Bob replaces the vacuum
input into the beamsplitter by one of his TMSV modes, and collects the output.
Correlations between the output and his retained TMSV mode provide him with an
additional advantage, while the attack manifests itself just as thermal noise in
the channel. Charlie measures an excess noise $\xi$ above shot noise.}
\label{fig:5}
\end{figure}

The thermal loss channel exhibits both loss and excess noise and can be modelled
by a beamsplitter with a thermal state $\rho_{th}\left(\bar{n}\right)$ input
into the fourth port, where $\bar{n}$ is the average number of photons in the
thermal state. However, the presence of this thermal noise will allow an
eavesdropping Bob to hide a more general attack, known as an \emph{entangling
cloner} attack \cite{Grosshans2003, Weedbrook2012}. In this attack, Bob starts
with an entangled two-mode squeezed vacuum (TMSV) state, and one of the two
entangled modes is injected into the fourth port of the beamsplitter,
Fig.~\ref{fig:5}. Once again he collects the reflected output, and performs an
optimal collective measurement on his two modes. Since the TMSV state purifies
the thermal state \cite{Weedbrook2012}, this attack manifests itself just as
thermal noise in the channel.

\par 
After the beamsplitter, Bob holds a two-mode state which is entangled with
Charlie's. The full state is
\begin{align}\label{eqn:state}
|\Psi\rangle_{B_1^\prime B_2 C}^j =
&\hat{D}_{B_1^\prime}\left(\sqrt{1-T}\alpha\right)
\hat{D}_C\left(\sqrt{T}\alpha\right) \notag \\
&\sum_{m=0}^\infty
\frac{G_m}{\sqrt{m!}}\left(\sqrt{T}\hat{a}^\dagger_{B_1^\prime} -
\sqrt{1-T}\hat{a}_C^\dagger\right)^m \notag \\
&|0\rangle_{B_1^\prime}|m\rangle_{B_2}|0\rangle_C
\end{align}
where $G_m = \left(\tanh r\right)^m/\cosh r$; $\bar{n} = \sinh^2 r$ is the
average number of thermal photons in one mode of the input TMSV state; and
$\hat{D}\left(\alpha\right) = \exp\left(\alpha\hat{a}^\dagger -
\alpha^*\hat{a}\right)$ is the displacement operator. After performing
heterodyne measurement on mode $C$ and receiving outcome $c \in \mathbb{C}$, the
state $\rho_{B|c}^j\left(\bar{n}\right) = \langle c
|\Psi\rangle\langle\Psi|c\rangle$ can be used to calculate the Holevo
information as before via Eqs.~(\ref{eqn:holevo}),~(\ref{eqn:int}). The states
$|\Psi\rangle_{B_1^\prime B_2 C}$ and $\rho_{B|c}^j\left(\bar{n}\right)$ are
derived in Appendix~\ref{appendix:tmsv}.
\par
Finally, we note that Charlie's probability of measuring $c \in \mathbb{C}$ when
Alice sends state $|\alpha_k\rangle$ through the channel with transmission $T$
and thermal noise input $\rho_{th}\left(\bar{n}\right)$ is
\cite{Papanastasiou2018}
\begin{equation}\label{eqn:probtherm}
p\left(c|\alpha_k\right) \left(\bar{n}\right) = \frac{\exp\left(-\frac{|c -
\sqrt{T}\alpha_k|^2}{1 + \left(1-T\right)\bar{n}}\right)}{\pi \left(1 +
\left(1-T\right) \bar{n}\right)}
\end{equation}
and so the excess noise measured at Charlie is $\xi =
\left(1-T\right)\bar{n}/2$. From Eq.~(\ref{eqn:probtherm}) we can calculate
$p_{err} \left(\xi\right)$ as the probability
$P\left(\re{c}<0|\alpha\right)\left(\bar{n}\right)$ that Charlie's heterodyne
output eliminates the sent state, analogously with Eq.~(\ref{eqn:perr}).

\subsection{Signature length $L$}
Now that we have calculated $p_e$ and $p_{err}$ for both beamsplitter and
entangling-cloner attacks, the probability $\varepsilon_{fail}$ that the
protocol fails can be found by calculating via
Eqs.~(\ref{eqn:erob}),~(\ref{eqn:erep}),~(\ref{eqn:eforg}) the probability that
the protocol is not robust $\varepsilon_{rob}$; the probability of successful
repudiation $\varepsilon_{rep}$; and the probability of successful forgery
$\varepsilon_{forg}$б. For a figure of merit, we assume that the protocol can
fail in any of these ways with equal probability and set
\begin{equation}
\varepsilon_{fail} = \varepsilon_{rob} = \varepsilon_{rep} = \varepsilon_{forg}.\end{equation}
By choosing $s_B = p_{err} + \left(p_e + p_{err}\right)/4$ and $s_C = p_{err} +
3 \left(p_e - p_{err}\right)/4$ we satisfy the second two equalities, and so the
overall probability of failure becomes
\begin{equation}\label{eqn:efail}
\varepsilon_{fail} \le 2 \exp\left(-\left(p_e - p_{err}\right)^2
\frac{L}{16}\right)
\end{equation}
provided that $p_e \ge s_C \ge s_B \ge p_{err}$. The security parameter $g = p_e
- p_{err}$ quantifies the advantage that an honest party holds over a dishonest
party, and if $g >0$ then our QDS protocol can be made arbitrarily secure by an
appropriate choice of $L$. The signature length $L$ required to sign $m$ to a
security level $\varepsilon_{fail}$ may thus be calculated using
Eq.~(\ref{eqn:efail}).

\section{Performance of the protocol}\label{sec:performance}
The main figure of merit for a QDS protocol is the signature length, $L$,
required to sign a $1$~bit message to a given security level
$\varepsilon_{fail}$. We choose the probability of failure to be
$\varepsilon_{fail} = 0.01\%$ and solve Eq.~(\ref{eqn:efail}) for $L$, under
both beamsplitter and entangling-cloner attacks, corresponding to $\xi = 0\%$
and $\xi > 0\%$, respectively, and for several different coherent state
amplitudes used in the alphabet.
\par
\begin{figure}[htp]
\centering
\includegraphics[width=\linewidth]{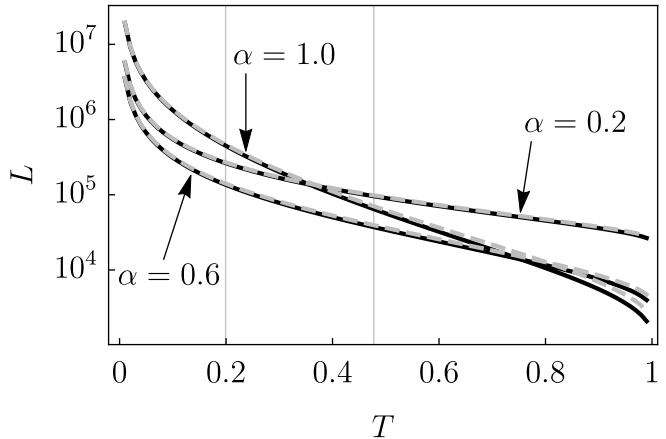}
\caption{\label{fig:6}Signature lengths $L$ required to securely sign a $1$~bit
message, for channel transmission $T$ and coherent state amplitude $\alpha$. The
length $L \rightarrow \infty$ as $T \rightarrow 0$, but remains modest at the
realistic distances denoted by vertical gridlines. Left gridline: $T = 0.20$
(approx. $20$~km fiber); right gridline: $T = 0.48$ (approx. $1$~km fiber).
Solid: $\xi = 0\%$ (beamsplitter attack). Dashed: $\xi = 2\%$ (entangling-cloner
attack). The required signature length $L$ is strongly influenced by the choice
of $\alpha$.}
\end{figure}

The signature lengths for the $\mathcal{A}_4$ alphabet are displayed in
Fig.~\ref{fig:6}. As expected, we see that $L\rightarrow \infty$ as $T
\rightarrow 0$ and the protocol can no longer be made secure in this limit.
However, for all $T>0$, the security parameter $g$ is positive and so the
protocol is secure, albeit with infeasibly large $L$ for the smallest values of
$T$. The presence of realistic amounts of excess noise increases $L$ at all $T$,
with increasingly drastic effects at small $T$--even though Charlie will allow
fewer thermal photons in the channel as $T$ decreases.
\par
For values of $T$ corresponding to realistic metropolitan distances we observe
that our QDS scheme can be made secure with surprisingly short signature lengths
$L$. Assuming optical fiber with $0.2$~dB loss per km, we calculate $T$
corresponding to $1$~km and $20$~km channels. At these distances, displayed in
the vertical gridlines of Fig.~\ref{fig:6}, we can securely sign a $1$~bit
message with only $L \sim O\left(10^5\right)$ coherent states. Combined with
fast sending rates typical to the CV platform, this opens up the possiblity of
signing a message in times competitive with the $O\left(10^{-1}\right)$~seconds
found in DV schemes (\cite{Collins2017}, and see Fig.~$7$ of
\cite{Collins2018}). For example, with a feasible sending rate of $100$~MHz our
protocol could securely sign a $1$~bit message in
$O\left(10^{-3}\right)$~seconds over $20$~km.
\par
The signature lengths required under our new security proof are shorter than
under our previous protocol, despite now making fewer assumptions about the
power of an eavesdropping party. For example, at $T = 0.5$, our current protocol
gives $L = 34139$ whereas the protocol from \cite{Croal2016} would give $L =
44010$. This improvement is beacuse in our new protocol we have chosen $\Phi_m^B
\ne \Phi_m^C$, so a dishonest party is forced to eavesdrop and thus receives an
imperfect copy of Charlie's states, whereas previously they were given a perfect
copy \cite{Croal2016, Amiri2016}.
\par
\begin{figure}[htp]
\centering
\includegraphics[width=\linewidth]{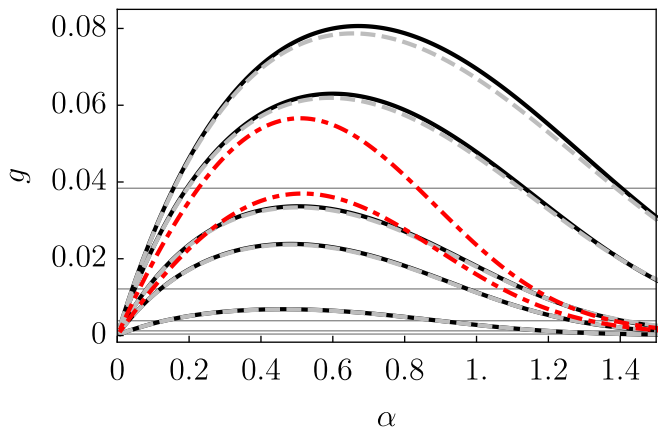}
\caption{(Color online). Security parameter $g$ as it varies with $\alpha$ for
$T = \{0.61\text{ (upper curve)}, 0.47, 0.19, 0.11, 0.01\text{ (lower
curve)}\}$. Solid: $\xi = 0\%$. Dashed: $\xi = 1\%$. The optimal $\alpha$ which
players should pick varies with $T$ but only slightly varies with $\xi$.
Horizontal gridlines denote $O\left(L\right)$ starting from $L \sim 10^5$ at $g
= 0.038$ (top) and increasing by a factor of $10$ at subsequent lower gridlines,
Eq.~(\ref{eqn:efail}). Red, dot-dashed: The $g$ varying with $\alpha$ for $1$~km
(upper curve) and $20$~km fiber (lower curve) under the previous protocol
\cite{Croal2016}, for $\xi = 0\%$. Note that under \cite{Croal2016} the optimal
$\alpha$'s do not vary with either $T$ or $\xi$.}
\label{fig:7}
\end{figure}

To understand the optimal behaviour of our protocol, we consider security
parameter $g$ instead of signature length, via Eq.~(\ref{eqn:efail}). We observe
in Fig.~\ref{fig:7} that the maximum $g$--therefore smallest $L$--varies
strongly with $T$ and $\alpha$ and only slightly with $\xi$. Therefore, for a
given channel it is important to pick the optimal $\alpha$ in order to minimise
the quantum resources required for security. This is in sharp contrast to
\cite{Croal2016}, shown in Fig~\ref{fig:7} by red, dot-dashed lines, where the
optimal $\alpha \approx 0.5$ for all channels.
\par
\begin{figure}[htp]
\centering
\includegraphics[width=\linewidth]{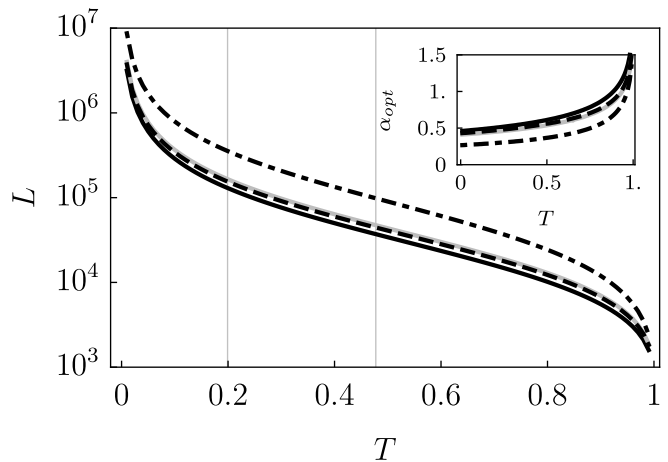}
\caption{Optimal signature length $L$ for $\xi = 0\%$. At each transmission $T$
the coherent state amplitude $\alpha_{opt}$ is chosen to minimize $L$. We have
considered alphabets $\mathcal{A}_4$, $\mathcal{A}_6$, $\mathcal{A}_8$, and
$\mathcal{A}_2$. Solid: $\mathcal{A}_4$. Dashed: $\mathcal{A}_6$. Gray, solid:
$\mathcal{A}_8$. Dot-dashed: $\mathcal{A}_2$. Inset: the corresponding
$\alpha_{opt}$. Choosing an alphabet size larger than $\mathcal{A}_4$ decreases
the optimal $\alpha_{opt}$ while slightly increasing the required signature
length $L$. As the alphabet size increases it becomes closer to a Gaussian
distribution, and so the beamsplitter and entangling-cloner attack become
increasingly optimal. The largest jump in protocol efficiency occurs from
$\mathcal{A}_2$ to $\mathcal{A}_4$. Vertical gridlines denote $T$ corresponding
to the $20$~km (left) and $1$~km (right) fiber.}
\label{fig:8}
\end{figure}

For each channel we minimize $L$ by optimizing over $\alpha$, and the results
are plotted in Fig.~\ref{fig:8}, with the required $\alpha_{opt}$ displayed in
the inset. For large $T$ a large $\alpha$ is optimal. In this case an
eavesdropper gains little information so honest parties should try to minimise
their mismatch rate. As $T$ decreases a smaller $\alpha$ will increase the
eavesdropper's mismatch rate but at the cost of also increasing honest
mismatches. Taking $\xi \ne 0\%$ was found to slightly decrease $\alpha_{opt}$.
\par
We have also considered the alphabet sizes $\mathcal{A}_6$, $\mathcal{A}_8$ and
$\mathcal{A}_2$, Appendix~\ref{appendix:larger}, with their optimal $L$'s also
plotted in Fig.~\ref{fig:8}. Surprisingly, although for larger alphabets the
optimal $\alpha$ is decreased, the minimal $L$ is slightly increased. As has
been found elsewhere \cite{Leverrier2011}, the biggest leap in behaviour should
occur between $\mathcal{A}_2 \rightarrow \mathcal{A}_4$, and indeed this is what
we see, noting that for $\mathcal{A}_2$ we no longer need to think about an
eliminated signature or ambiguity and we simply consider optimal guessing
probabilities. As the alphabet tends towards a Gaussian mixture of coherent
states, we expect the two attack strategies considered in this paper to become
increasingly optimal, which explains the slight increase in $L$ for larger
alphabets.

\section{Discussion}\label{sec:discussion}
Quantum digital signatures, which allow for secure authentication of a classical
message, have only recently been proven secure against a quantum eavesdropper on
the channels \cite{Amiri2016, Puthoor2016, Yin2016}. In the present paper, we
have advanced QDS protocols operating on the continuous-variable platform by
providing security against beamsplitter and entangling-cloner attacks on the
quantum channels. Surprisingly, short signature lengths (even shorter than in
\cite{Croal2016} under the assumption of secure quantum channels) are sufficient
to ensure secure QDS over metropolitan distances. The conceptually new security
proof has enabled us to include the fact that for each eliminated signature
element there are multiple ``correct'' declarations which a dishonest player can
make, and which must be taken into account.
\par
Our security proof relied on several assumptions which reflect the
state-of-the-art of CV quantum cryptography with our chosen alphabet, but which
future work should endeavour to relax \cite{Papanastasiou2018, Zhao2009,
Bradler2018, Weedbrook2012}. Firstly, the eavesdropping attacks permitted by a
malevolent Bob in this work do not give him the full power afforded by quantum
mechanics. The non-Gaussianity of our alphabet is restrictive here, and the
entangling-cloner attack is only expected to be optimal as the limiting case
that the alphabet becomes Gaussian, i.e. for $\alpha \rightarrow 0$
\cite{Navascues2006, Garcia-Patron2006}. For our alphabet with discrete
modulation, a wider class of non-Gaussian attacks may provide an eavesdropper
with an advantage, and more work is needed to explore optimal classes of
non-Gaussian attack. In a general protocol these effects could be taken into
account for example by tomographically reconstructing the state $\rho_B$ which
maximizes the Holevo information $\chi$, while remaining consistent with
Charlie's measurement outcomes--though we note that the resources required for
this will be expensive and may undermine the ease-of-use which our scheme
currently boasts. Another possible route towards improving security would be an
extension of results known for QKD with $2$~state \cite{Zhao2009} and $3$~state
\cite{Bradler2018} alphabets to our $\mathcal{A}_4$, noting recent progress in
\cite{Papanastasiou2018}.
\par
Techniques used in our security proof will in future allow us for possibility to
explore different security tasks, such as secret sharing \cite{Kogias2017} or
oblivious transfer \cite{Broadbent2015}, and to design protocols relying on the
same modest physical requirements which we used here. One may also begin to
consider consider finite-size effects \cite{Tomamichel2016}, which are intrinsic
to any QDS scheme, noting the operational links between the guessing
probabilities considered in this paper and the smooth min-entropy
\cite{Konig2009}. Advances in calculating optimal lower bounds for the smooth
min-entropy will have immediate and direct application to CV QDS, and may be
readily incorporated into our security proof.
\par
The security of our QDS protocol and the short time required to sign a message,
stemming both from the conceptually new security proof and the practical
advantages of the CV platform, make CV QDS an attractive scheme for secure
communications in a quantum future. It may soon be possible to move to
real-world implementation of our scheme, with opportunity to run alongside
related QKD schemes \cite{Heim2014}.

\bibliographystyle{apsrev4-1}
\bibliography{bib-qds}

\appendix
\section{Hoeffding's inequalities}\label{appendix:hoeffding}
Hoeffding's inequalities \cite{Hoeffding1963,Watrous2018} provide a bound for
the probability that the empirical mean of $n$ independent outcomes differs from
the expectation. The most useful form for our purposes is shown below, and we
briefly demonstrate how they may be utilised in our security proof. A full
treatment can be found in \cite{Croal2016, Collins2014}.
\par
\subsection{Hoeffding's inequalities.} Let $X_1,\ldots, X_n$ be independent
binary random variables. Let $\bar{X}$ be their empirical mean, and
$\mathbb{E}\left(\bar{X}\right)$ their expected value. Then for every
$\varepsilon \ge 0$ the following are true
\begin{align}\label{eqn:hoeffding}
P\left(\bar{X} - \mathbb{E}\left(\bar{X}\right) \ge \varepsilon \right) &\le
\exp\left(-2 \varepsilon^2 n\right) \notag \\
P\left(\mathbb{E}\left(\bar{X}\right) - \bar{X} \ge \varepsilon \right) &\le
\exp\left(-2 \varepsilon^2 n\right)
\end{align}
\par
\subsection{Application to QDS.} Let $\mathcal{F}$ be a string of declared
phases, and $\mathcal{G}$ be an eliminated signature. Define a string $E$
\begin{equation*}\label{eqn:error}
E_j = 
\begin{cases}
1 & \text{if $F_j$ is eliminated in $G_j$} \\
0 & \text{otherwise}
\end{cases}
\end{equation*}
which measures the number of mismatches between $\mathcal{F}$ and $\mathcal{G}$.
All strings are of length $n$. We wish to bound the probability that the number
of mismatches is below some threshold $s n$, or equivalently the probability
$P\left(\bar{E} \le s\right)$ that the observed mismatch rate $\bar{E} = 1/n
\sum_{j=1}^n E_j$ is below $s$. Then we have
\begin{align}\label{eqn:ehoeff}
P\left(\bar{E} \le s\right) &= P\left(\mathbb{E}\left(\bar{E}\right) - \bar{E}
\ge \mathbb{E}\left(\bar{E}\right) - s\right) \notag \\
&\le \exp\left(-2\left(\mathbb{E}\left(\bar{E}\right) - s\right)^2 n\right)
\end{align}
where the equality follows trivially provided that $\mathbb{E}\left(E\right) - s
\ge 0$, and the inequality is an application of Eq.~(\ref{eqn:hoeffding}).
Bounds on $P\left(s \le \bar{E}\right)$ may be similarly derived:
\begin{align}\label{eqn:ehoeff2}
P\left(s \le \bar{E}\right) \le \exp\left(-2\left(s -
\mathbb{E}\left(\bar{E}\right)\right)^2n\right).
\end{align}
\section{Larger alphabets}\label{appendix:larger}
We show that our central result, Eq.~(\ref{eqn:hpe}), holds for all alphabets
$\mathcal{A}_N$ with $N = 2 k$; $k \in \mathbb{N}$; consisting of coherent
states equally distributed about the origin in phase space. We also remark on
any required modifications to the calculations presented in the paper.
\par
During the protocol, Bob and Charlie should eliminate exactly $N/2$ coherent
states, using the same strategy as in Fig~\ref{fig:2}. Otherwise the running of
the protocol remains the same.
\par
As before, Eqs.~(\ref{eqn:chain1}),~(\ref{eqn:chain2}),~(\ref{eqn:chain3}), we
start with $H\left(E_j, x_1^j,\dots,x_{N/2}^j|y_j\right)$ and use the chain rule
for conditional entropies twice, giving
\begin{equation*}
H\left(x_1^j,\dots,x_{N/2}^j|y_j\right) = H\left(x_1^j,\dots, x_{N/2}^j |
E_j,y_j\right) + H\left(X_j|y_j\right)
\end{equation*}
once we have taken into account that $H\left(E_j |
x_1^j,\dots,x_{N/2}^j,y_j\right) = 0$. Using $H\left(E_j | y_j\right) \le
h\left(p_e\right)$ and the fact that Bob and Charlie eliminate exactly $N/2$ out
of $N$ possible alphabet states, we arrive at
\begin{equation*}
H\left(x_1^j,\dots,x_{N/2}^j|y_j\right) \le H\left(x_1^j,\dots, x_{N/2}^j |
E_j=0,y_j\right) + h\left(p_e\right),
\end{equation*}
therefore
\begin{align*}
&H\left(x_1^j,\dots,x_{N/2}^j\right) - \chi\left(x_1^j,\dots,x_{N/2}^j :
y_j\right) \notag \\
&\le H\left(x_1^j,\dots, x_{N/2}^j | E_j=0,y_j\right) + h\left(p_e\right).
\end{align*}
To complete the proof of Eq.~(\ref{eqn:hpe}) we simply observe
\begin{align*}
H\left(x_1^j,\dots,x_{N/2}^j\right) &= \log_2\left(N \times \frac{N}{2}!\right)
\notag \\
H\left(x_1^j,\dots, x_{N/2}^j | E_j=0,y_j\right) &= \log_2\left(\frac{N}{2}
\times \frac{N}{2}!\right),
\end{align*}
where we have taken into account relabeling, and Eq.~(\ref{eqn:hpe}) follows
immediately.
\par
The quantities used to calculate the Holevo information must also be altered in
order to reflect the different alphabet. Bob's a priori state becomes
\begin{equation}
\rho_B^j = \frac{1}{N}
\sum_{k=1}^N|\sqrt{1-T}\alpha_k\rangle_B\langle\sqrt{1-T}\alpha_k|,
\end{equation}
the state $\rho_{B|c}^j$ is similarly transformed--both for beamsplitter attack
and entangling-cloner attack. The integration limits of Eq.~(\ref{eqn:int}) are
also altered so that each segment now occupies an angular width of $2 \pi/N$.
Finally, since Bob and Charlie eliminate exactly $N/2$ of the alphabet, the
probability $p_{err}$ that a heterodyne measurement should eliminate Alice's
sent state remains unchanged.

\section{Two-mode squeezed vacuum}\label{appendix:tmsv}
We will calculate the $|\Psi\rangle_{B_1\prime B_2 C}^j$ and $\rho_{B|c}^j$
required for the entangling cloner attack, Fig.~\ref{fig:5}. Our starting point
is the state shared between Alice and Bob before the channel. Alice generates
coherent state $|\alpha\rangle$ and Bob generates a two-mode squeezed vacuum
(TMSV) state \cite{Leonhardt2010}. Then Alice and Bob share the three-mode state
\begin{equation}\label{eqn:beforechannel}
|\alpha\rangle_A |\text{TMSV}\rangle_{B_1 B_2} = \hat{D}_A\left(\alpha\right)
|0\rangle_A \sum_{m=0}^\infty G_m
\frac{\left(\hat{a}_{B_1}^\dagger\right)^m}{\sqrt{m!}}
|0\rangle_{B_1}|m\rangle_{B_2}
\end{equation}
where $|\alpha\rangle = \hat{D}\left(\alpha\right)|0\rangle$ and
$\hat{D}\left(\alpha\right) = \exp\left(\alpha \hat{a}^\dagger - \alpha^*
\hat{a}\right)$ is the displacement operator; and where we have written
$|m\rangle_{B_1} = \left(\hat{a}_{B_1}^\dagger\right)^m / \sqrt{m!}
|0\rangle_{B_1}$. The coefficient $G_m = \left(\tanh r\right)^m / \cosh r$ where
$r$ parametrises the number of thermal photons $\bar{n}$ in each of the two
modes via $\bar{n} = \sinh^2 r$.
\par
The beamsplitter transforms our creation operators as
\begin{equation}\label{eqn:beamsplitter}
\begin{pmatrix}
\hat{a}_C \\
\hat{a}_{B_1^\prime}
\end{pmatrix}
=
\begin{pmatrix}
\sqrt{T} & -\sqrt{1-T} \\
\sqrt{1-T} & \sqrt{T}
\end{pmatrix}
\begin{pmatrix}
\hat{a}_A \\
\hat{a}_{B_1}
\end{pmatrix}.
\end{equation}
Using Eq.~(\ref{eqn:beamsplitter}) we transform our input state to give
\begin{align}\label{eqn:afterchannel}
|\Psi\rangle_{B_1^\prime B_2 C}^j &=
\hat{D}_{B_1^\prime}\left(\sqrt{1-T}\alpha\right)\hat{D}_{C}\left(\sqrt{T}\alpha\right)
\notag \\
&\sum_{m=0}^\infty G_m \frac{\left(\sqrt{T}\hat{a}_{B_1^\prime}^\dagger -
\sqrt{1-T}\hat{a}_C^\dagger\right)^m}{\sqrt{m!}} |0\rangle_{B_1^\prime}
|m\rangle_{B_2}|0\rangle_C
\end{align}
where we have used the fact that $\hat{a}_{B_1^\prime}$ and $\hat{a}_C$ commute.
The state Eq.~(\ref{eqn:afterchannel}) may be computed by using the binomial
expansion on the brackets, and the state $\rho_{B|c}^j\left(\bar{n}\right) =
\langle c|\Psi\rangle\langle \Psi|c\rangle$ can now be computed.

\end{document}